\documentstyle[aps,prb,multicol,epsfig]{revtex}

\draft

\renewcommand{\narrowtext}{\begin{multicols}{2} \global\columnwidth20.5pc}
\renewcommand{\widetext}{\end{multicols} \global\columnwidth42.5pc}

\begin{document}

\title{Lanczos exact diagonalization study of field-induced
 phase transition for Ising and Heisenberg antiferromagnets}

\author{Seung-Pyo Hong, Choong-Seok Kim, Sung-Sik Lee and Sung-Ho Suck Salk}

\address{Department of Physics,
 Pohang University of Science and Technology,
 Pohang 790-784, Korea}


\maketitle

\vspace*{5mm}

\begin{abstract}
Using an exact diagonalization treatment of
Ising and Heisenberg model Hamiltonians,
we study field-induced phase transition 
for two-dimensional antiferromagnets.
For the system of Ising antiferromagnet
the predicted field-induced phase transition is of first order,
while for the system of Heisenberg antiferromagnet
it is the second-order transition.
We find from the exact diagonalization calculations
that the second-order phase transition (metamagnetism)
occurs through a spin-flop process
as an intermediate step.
\end{abstract}


\vspace*{5mm}

\narrowtext

\section{INTRODUCTION}
Much earlier (in 1932) the antiferromagnetic order was proposed
by N\'{e}el\cite{NEEL} in order to explain the low temperature behavior
of the magnetic susceptibility of certain metals.
Currently physical properties of low-dimensional quantum antiferromagnets
at low temperature are actively pursued.
The exact solution via Bethe Ansatz\cite{BETE} is limited to 
one-dimensional integral systems, but not extendible
to multidimensional systems.
Recently, one among the interesting subjects is
a numerical study of field-induced phase transition, namely,
the antiferromagnetic to ferromagnetic transition
in the two-dimensional systems of antiferromagnetically correlated electrons
under externally applied magnetic field.
By applying the dynamical mean field theory (DMFT) to the Hubbard model,
Held {\it et al.}\cite{HELD} studied
the microscopic origin of metamagnetism along an easy axis 
in antiferromagnets under external magnetic field.
Their study showed that at half filling a metamagnetic phase transition
arises via the first-order phase transition at low temperature
and that the second-order phase transition occurs 
near the N\'{e}el temperature.
On the other hand,
Bagehorn and Hetzel\cite{BAGE} observed 
the second-order phase transition at zero temperature 
from the projector quantum Monte Carlo (PQMC) calculation of the Hubbard model
with an easy axis.
Earlier using a Landau theory of free energy
Moriya and Usami\cite{MORI} revealed 
the second-order phase transition at low temperature
involving the mixed phase of ferro- and antiferromagnetic states.
However, Bagehorn and Hetzel questioned whether the mixed phase would survive
at a small region of magnetic field
if exact electron correlations were implemented. 
In the present study,
by performing the exact diagonalization calculations of the
two-dimensional systems of antiferromagnetically correlated electrons
under applied magnetic field,
we examine the nature of field-induced phase transition
(metamagnetism) at 0 K
for both Ising and Heisenberg antiferromagnets.

\section{LANCZOS EXACT DIAGONALIZATION CALCULATIONS OF 
         FIELD-INDUCED PHASE TRANSITIONS}
With the inclusion of Zeeman coupling term,
the two-dimensional $t$-$J$ model Hamiltonian is written,
\begin{eqnarray}
H &=& - t\sum _{\langle ij \rangle, \sigma}  
\Bigl( (1-n_{i,-\sigma})c_{i \sigma}^{\dagger}c_{j \sigma}(1-n_{j,-\sigma}) 
+ {\rm c.c.} \Bigr )
 \nonumber \\
&&
+ J \sum _{\langle ij \rangle}({\bf S}_i \cdot {\bf S}_j -\frac{1}{4}n_i n_j)
-  h\sum_{i \sigma} \sigma c_{i\sigma}^\dagger c_{i\sigma}^{} ~,
\label{tjmodel}
\end{eqnarray}
where $c_{i \sigma}$ $(c_{i \sigma}^{\dagger})$ 
is the electron annihilation (creation) operator,
${\bf S}_i$ is the electron spin operator at site $i$,
and $h$ is the Zeeman energy, $\mu_{\scriptscriptstyle B}^{} B$.
Here $\mu_{\scriptscriptstyle B}^{}$ is the Bohr magneton
and $B$ is the strength of external magnetic field. 
$\sigma=+1~(-1)$ for up spin (down spin).
In the case of half filling, 
the $t$-$J$ model Hamiltonian effectively reduces to 
the Heisenberg model Hamiltonian.

The above Hamiltonian in Eq.~(\ref{tjmodel}) will be diagonalized by the
Lanczos exact diagonalization method.\cite{DAGO94}
For the exact diagonalization treatment of antiferromagnetic correlations
uniform and staggered magnetizations are defined by,\cite{DAGO}
\begin{equation}
\label{b}
 \langle ( m_q^\ell )^2 \rangle = 
 \left< \left( \frac{1}{N} \sum_i e^{i {\bf q} \cdot {\bf r}_i} 
 S^\ell_i \right)^2 \right> ~,
\end{equation}
where $\ell=x, y, z$. 
Here $(m_q^\ell)^2$ represents the square of
the uniform and the staggered magnetizations
in the $\ell$th direction corresponding to
$q=(0,0)$ and $q=Q\equiv(\pi,\pi)$ respectively.
$N$ is the total number of lattice sites. 
We calculate the variation of the uniform and the staggered magnetizations
with vertically applied magnetic field
to a $4\times 4$ square lattice 
of antiferromagnetically correlated electrons
by satisfying periodic boundary conditions.
For the parameters we choose $t=1$ and $J=0.4$.
Lanczos iteration is terminated 
when the ground state energy is converged
within the error bound of $10^{-10}$.

\begin{figure}
\centering
\epsfig{figure=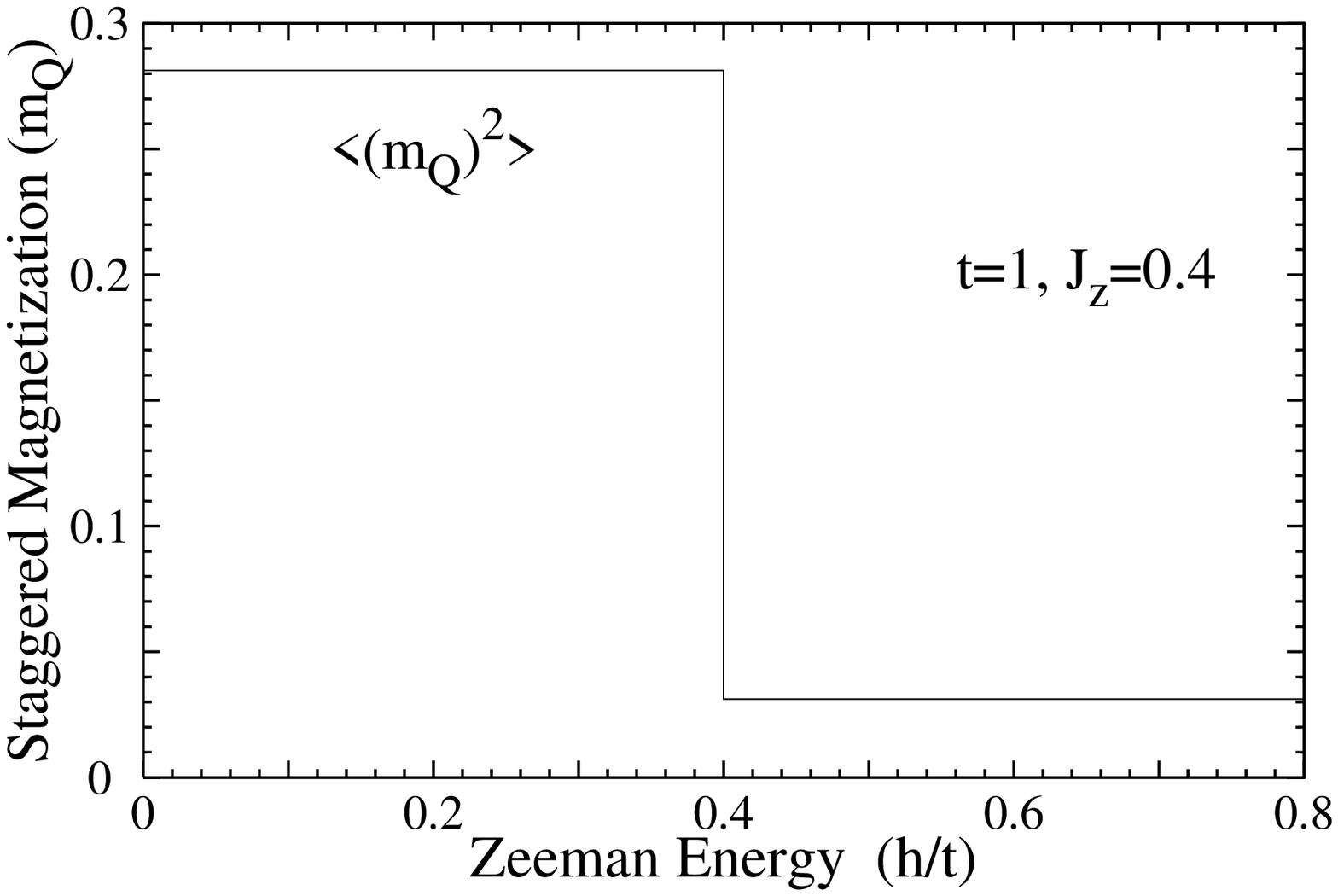, width=8cm} \\
(a) \\[5mm]
\epsfig{figure=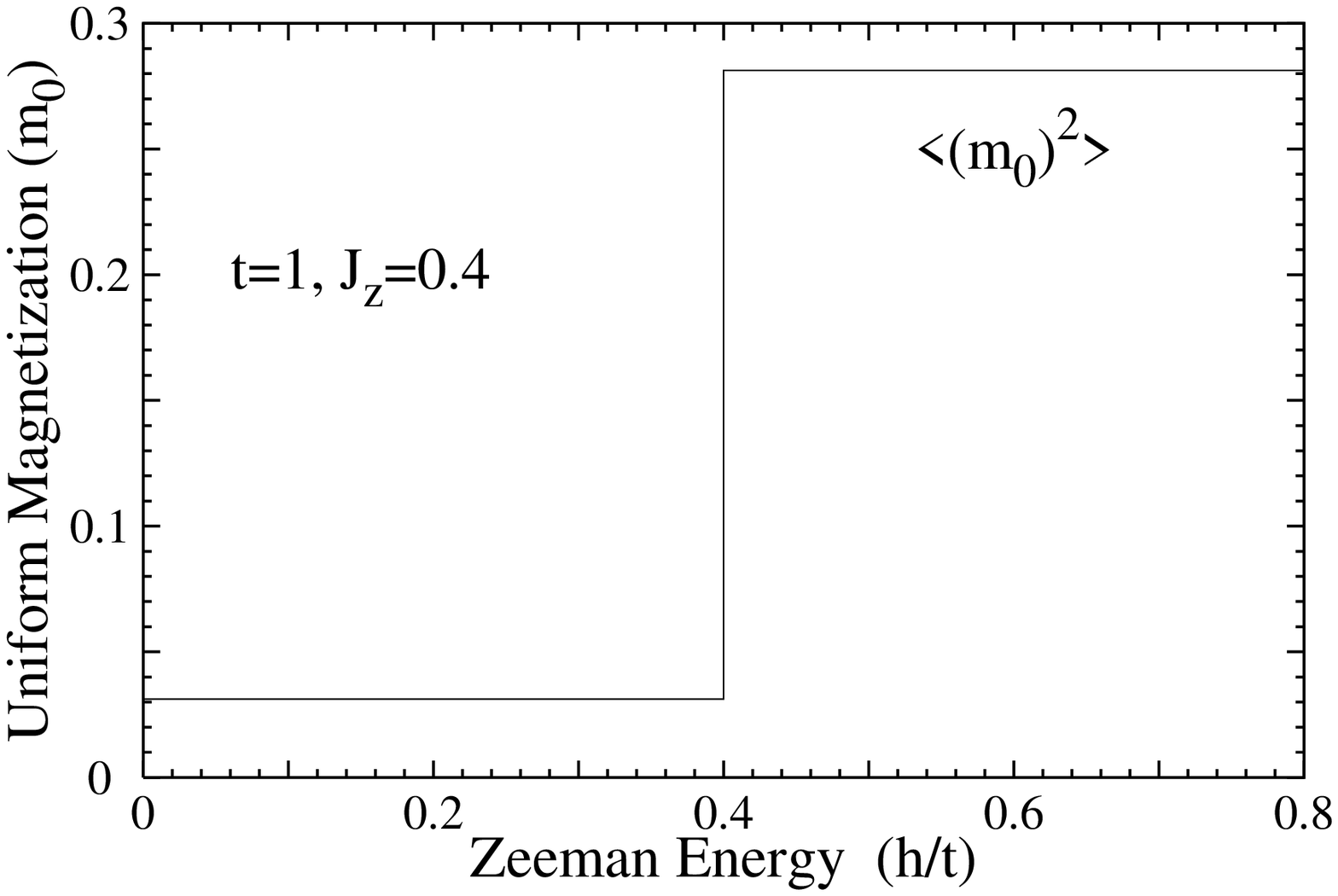, width=8cm} \\
(b) \\[5mm]
\caption{
 Staggered and uniform magnetizations
 for the two-dimensional Ising system
 as a function of Zeeman energy.
 (a) staggered magnetization $m_{\scriptscriptstyle Q}^{}$
 and (b) uniform magnetization $m_0^{}$.}
\label{fig:ising}
\end{figure}

For the Ising systems, $H = J \sum_{\langle ij\rangle} S_i^z S_j^z$,
we find that the metamagnetic phase transition is of first order,
as is shown in Fig.~\ref{fig:ising}.
In our previous work,\cite{KIM} we calculated 
the staggered and the uniform magnetizations for the Ising system
by using Hubbard model Hamiltonian in mean-field approximation.
The first-order phase transition was also predicted.
In both approaches we note that
there exists no mixed phase 
(of both $m_{\scriptscriptstyle Q}^{} \neq 0$ and $m_0^{} \neq 0$).
In other words, discontinuity from staggered to uniform magnetization 
is observed at a critical magnetic field, say,
$h=0.4t$ in Fig.~\ref{fig:ising}
(the staggered magnetization beyond the critical field
and the uniform magnetization below the same point
did not completely vanish owing to finite-size effect).
Thus in the Ising systems metamagnetism occurs 
as a first-order phase transition.
This is in agreement with the DMFT study of Held {\it et al}.\cite{HELD}
On the other hand, Bagehorn and Hetzel showed from their PQMC calculation that 
metamagnetism occurs via a second-order phase transition.

\vspace*{7mm}

\begin{figure}
\centering
\epsfig{figure=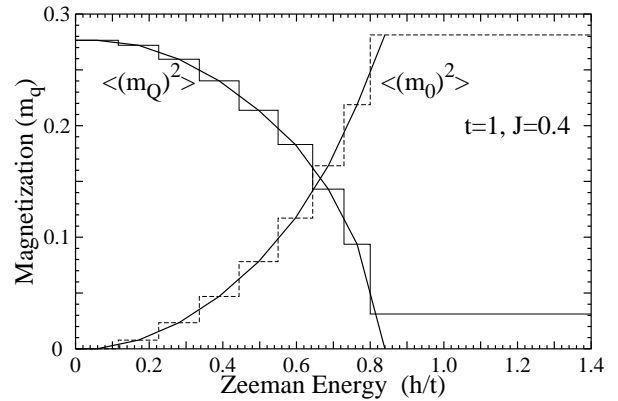, width=8cm} \\[5mm]
\caption{
 Staggered $m_{\scriptscriptstyle Q}^{}$ (solid line)
 and uniform $m_0^{}$ (dotted line) magnetizations
 for the two-dimensional Heisenberg system
 as a function of Zeeman energy.
 Solid curves are guide lines for the eyes.}
\label{fig:HmagQ0}
\end{figure}

For the Heisenberg systems,
we show the predicted magnetization as a function of applied field
in Fig.~\ref{fig:HmagQ0}.
The stepwise curves are inevitably formed
owing to the size effect.
As the number of lattice sites increases,
the stepwise curves are expected to turn smooth.
The solid curves in the figure are interpolated ones 
only for the visual aid.
The second-order field-induced transition is observed for metamagnetism at 0 K.
The magnetic susceptibility, $\chi \,(=\! \frac{dm}{dh})$ changes 
from a negative to a positive value
at a crossing point 
between the lines of the staggered magnetization (solid line)
and of the uniform magnetization (dotted line).
This indicates the occurrence of a mixed phase
of the ferro- and antiferromagnetic phases
during the process of metamagnetism.
The predicted antiferromagnetic susceptibility vanishes 
beyond a `critical field' corresponding to the Zeeman energy, $h \simeq 0.8t$
(the staggered magnetization 
did not completely vanish owing to the size effect).
The uniform magnetization is predicted to remain constant 
beyond the same point.
Although not to be directly compared,
due to differences in the space dimension of antiferromagnet,
such a tendency of uniform magnetization 
was experimentally observed 
for the three-dimensional crystal of ${\rm Y}({\rm Co}_{1-x} {\rm Al}_x)_2$ 
at low temperature
by Goto {\it et al}.\cite{GOTO} 

\begin{figure}
\centering
\epsfig{figure=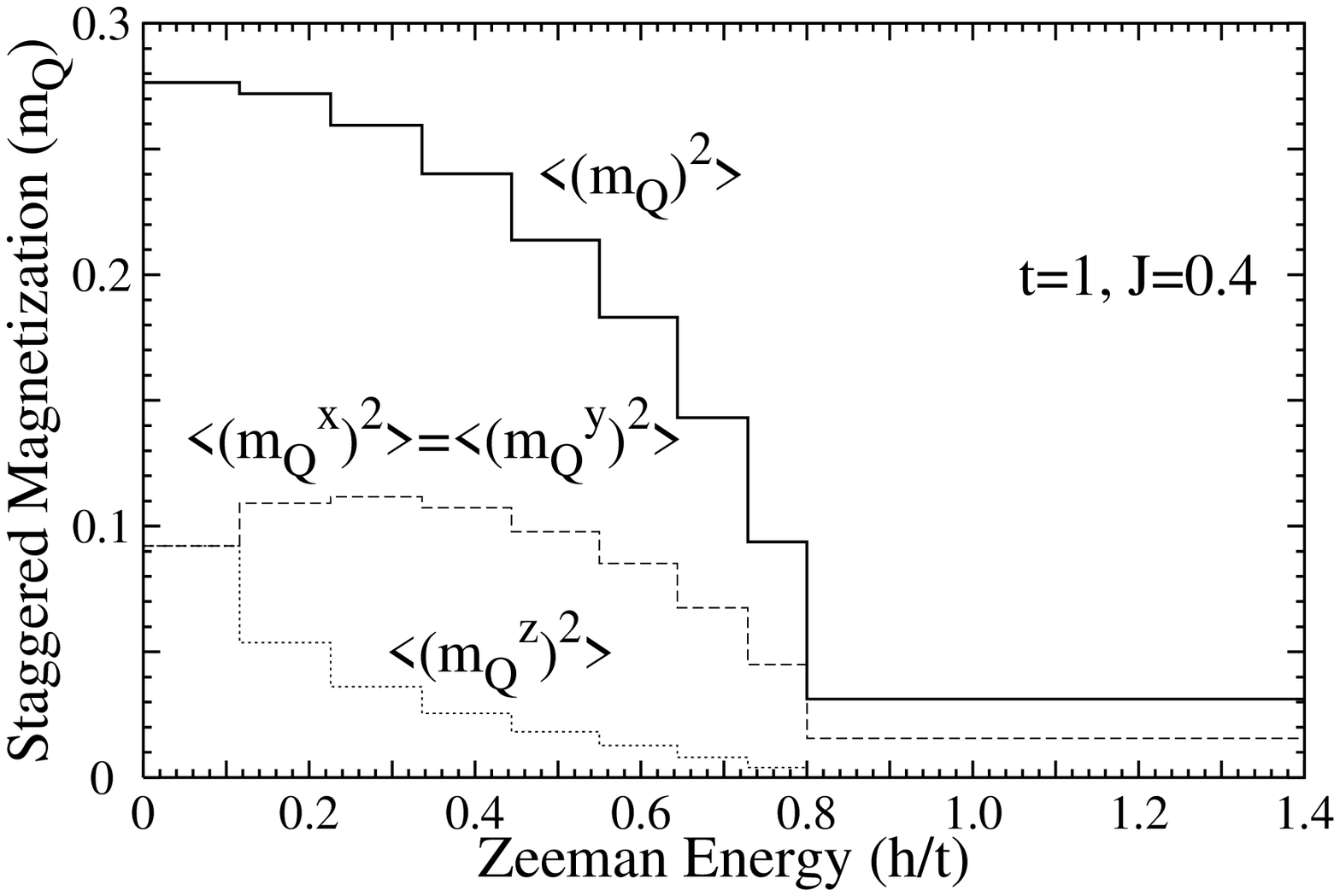, width=8cm} \\
(a) \\[2mm]
\epsfig{figure=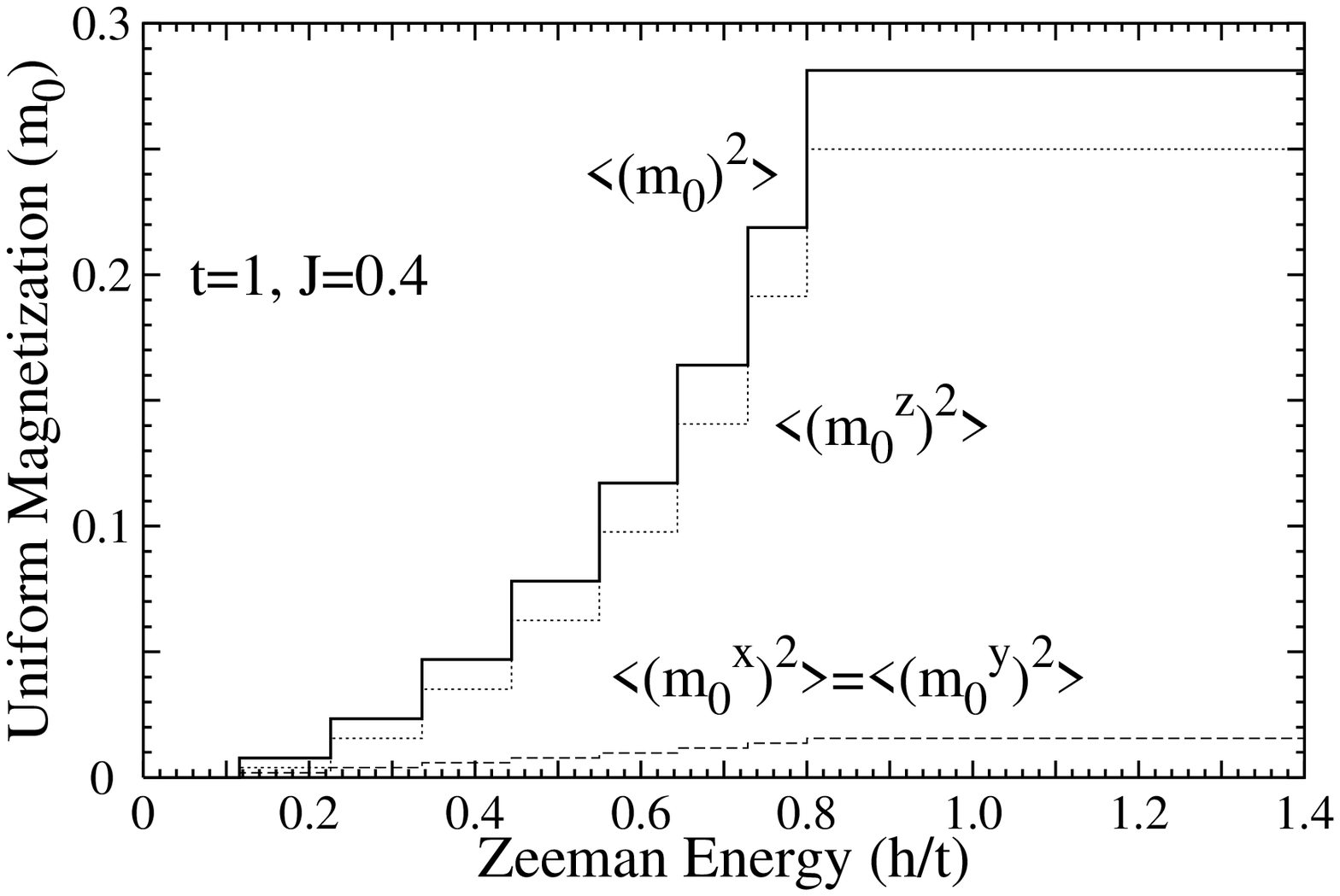, width=8cm} \\
(b) \\[3mm]
\caption{
 Staggered and uniform magnetizations
 for the two-dimensional Heisenberg system
 as a function of Zeeman energy.
 (a) Decomposition of staggered magnetization ($m_{\scriptscriptstyle Q}^{}$)
 and (b) decomposition of uniform magnetization ($m_0^{}$)
 into directions parallel ($m_q^x$ and $m_q^y$)
 and perpendicular ($m_q^z$) to the plane respectively for $q=Q,0$.}
\label{fig:heisenberg}
\end{figure}

\begin{figure}
\centering
\epsfig{figure=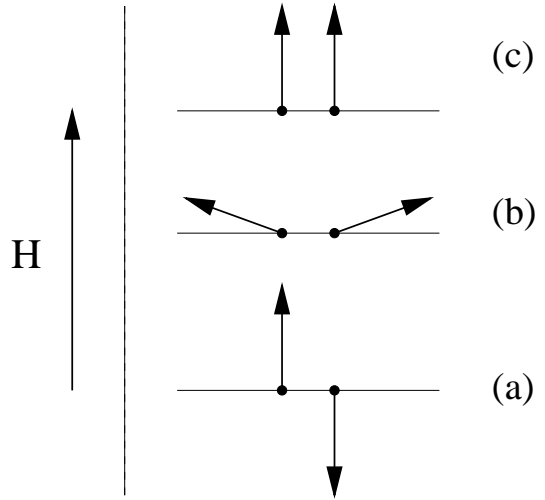, width=7cm} \\[5mm]
\caption{
 Spin flopping and metamagnetism under applied magnetic field.
 The magnetic field is applied in the $z$-direction.
 The horizontal solid line represents the two-dimensional $x$-$y$ plane.
 (a) antiferromagnetic spin order, (b) spin flop,
 and (c) ferromagnetic spin order.}
\label{fig:spin}
\end{figure}

In order to find the cause of the second-order phase transition
with the Heisenberg antiferromagnet,
in Fig.~\ref{fig:heisenberg} 
we show both the staggered and uniform magnetizations
in the direction parallel to
and perpendicular to the plane respectively.
As the external magnetic field increases, 
the $x$- and $y$-components of the staggered magnetization 
are predicted to increase particularly in the region of low field,
$h$ \raisebox{-0.3mm}{$\stackrel{\scriptstyle <}{\scriptstyle \sim}$} $0.3t$
in Fig.~\ref{fig:heisenberg}(a),
whereas the $z$-component decreases in the same region.
The antiferromagnetic order projected onto the plane
(perpendicular to the external magnetic field) 
tends to persist to a point,
while its $z$-component $m_{\scriptscriptstyle Q}^z$
monotonically decreases and
vanishes at $h \simeq 0.8t$.
This can be further explained as follows.
In the absence of magnetic field, 
spins align into a strong antiferromagnetic state 
due to the Heisenberg interaction 
as is shown in Fig.~\ref{fig:spin}(a).
As the magnetic field in the $z$-direction increases,
spin-flop process occurs as an intermediate state 
by exhibiting the components of 
antiferromagnetic spin alignment on the $x$-$y$ plane, 
as depicted in Fig.~\ref{fig:spin}(b).
For the Heisenberg antiferromagnet, the spin flop, that is,
the $x$- and $y$-components of the staggered magnetization peak
at a particular value of the applied magnetic field,
say, $h \simeq 0.3t$ as is shown in Fig.~\ref{fig:heisenberg}(a).
As the external magnetic field further increases, 
the Zeeman effect begins to dominate the Heisenberg interaction
with disappearance of the spin-flop process
by allowing ferromagnetic configurations,
as is shown in Fig.~\ref{fig:spin}(c). 
On the other hand, for the Ising system
the first-order phase transition was predicted
with no involvement of spin-flop process before the metamagnetic transition,
as is shown in Fig.~\ref{fig:ising}.
Thus we claim that
antiferromagnetic spin interactions involving $x$- and $y$-components
are responsible for the second-order phase transition 
accompanying a spin-flop process as an intermediate step.

\section{CONCLUSION}
By applying the exact diagonalization method to Heisenberg antiferromagnets 
under external magnetic field,
we examined how the second-order phase transition occurs 
from an antiferromagnetic to a ferromagnetic state 
at a critical magnetic field.
The second-order phase transition occurred in the presence of
the mixed phase near the critical magnetic field.
According to the present exact diagonalization study,
the second-order phase transition was induced
through the presence of spin-flop process
owing to the influence of
the $x$- and $y$-components of antiferromagnetic spin interactions.
In short we observed that
for the case of two-dimensional Ising systems
the field-induced transition from the antiferromagnetic state
to the ferromagnetic state is of first order
with no involvement of spin-flop process,
while for the case of two-dimensional Heisenberg antiferromagnetic systems
it is of second order
with the involvement of the spin-flop process as an intermediate process.

\acknowledgments{
S.-H.S.S. greatly acknowledges 
Korean Ministry of Education (BSRI-1998)
and the Center for Molecular Science 
at Korea Advanced Institute of Science and Technology for financial supports.}

\widetext

\end{document}